# Probabilistic Identification of Technology Tipping Points in Deeply Decarbonised Energy Systems

Gian Müller[1,2], Thomas Schöb[1], Jann M. Weinand[1], Iain Staffell[3]

[1]Institute of Climate and Energy Systems – Jülich Systems Analysis, Forschungszentrum Jülich GmbH, Jülich, 52425, Germany.

[2]Chair for Fuel Cells, RWTH Aachen University, Aachen, 52062, Germany

[3]Centre for Environmental Policy, Imperial College London, London, UK.

Corresponding Author: Iain Staffell, i.staffell@imperial.ac.uk

## Abstract

Energy policy is often guided by a small set of least-cost pathways to net-zero emissions, despite wide uncertainty in technology performance, fuel prices, demand and weather. To avoid overstating confidence in any single pathway, we quantify the likelihood of alternative technology pathways and identify the assumptions driving divergence, including the conditions under which technologies reach critical tipping points in competitiveness. We couple a sector-linked national optimisation model with Monte Carlo sampling (10,000 runs) across two European power systems (Germany and Great Britain) to generate probability distributions of capacity expansion and robust cost thresholds for key technologies. Results reveal substantial ambiguity in the future roles of wind versus solar, gas with carbon capture, and negative-emissions options. Tipping points vary widely with system conditions, while cross-country differences highlight the role of institutional constraints and resource endowments. Britain exhibits an "either-or" decision around nuclear power, investing if costs in 2035 fall below €4700/kW, otherwise favouring offshore wind. Germany's uncertainty centres on dispatchable low carbon options: gas with carbon capture (<€2100/kW), biomass with carbon capture (<€4200/kW) or hydrogen if electrolysis is below €560/kW. We reframe scenario analysis as risk management by linking uncertainty to cost targets and minimum deployment requirements for robust net-zero strategies.

# Introduction

Many countries have adopted increasingly ambitious climate targets and are implementing policies aimed at accelerating the transformation of energy systems towards low-carbon, sustainable production and consumption [1]. The design and evaluation of such policies are frequently informed by energy system studies, which commonly rely on optimisation-based modelling frameworks to derive cost-efficient transformation pathways [2].

Historically, most energy system optimisation models used to inform policy and decision-making have relied on deterministic modelling approaches [3,4]. In such frameworks, parameter uncertainty is typically addressed in a limited manner (see Supplementary Note 1). Studies commonly use single-factor sensitivity analyses, or a small number of qualitatively defined scenarios which are typically constructed from subjective assumptions regarding future developments [5]. This issue becomes particularly critical in the context of emerging and rapidly evolving technologies that are expected to play a key role in the decarbonisation of future energy systems. Technologies such as carbon capture and storage, power-to-X and direct air capture are frequently incorporated into energy system models despite being characterised by substantial techno-economic uncertainty, often associated with early stages of technological maturity and limited empirical deployment data (see Supplementary Note 2). As a result, the assumption of fixed techno-economic parameters – such as investment costs, efficiencies, or learning rates – can significantly influence the model's optimal system design and may lead to substantially different policy recommendations. Future electricity demand has become another highly uncertain dimension, with the potential for rapid growth contingent on the electrification of transport and heating, and growth of data centres and artificial intelligence (see Supplementary Note 3). A systematic assessment of parameter uncertainty in energy system models is therefore essential.

Explicitly addressing uncertainty through probabilistic modelling approaches enables a range of new analytical insights. For instance, it allows the assessment of the probability of technology deployment over time, the evaluation of the robustness of infrastructure expansion pathways, and the identification of key parameters that most strongly influence system outcomes. In addition, probabilistic analyses can reveal the diversity of possible transformation trajectories that satisfy climate targets, thereby highlighting the range of feasible policy and technology choices available during the energy transition. Such insights are particularly valuable for decision-makers as they provide a more nuanced understanding of risks, trade-offs, and the robustness of decarbonisation strategies. Frameworks for propagating uncertainty through complex models like Monte-Carlo techniques have rarely been used with national-scale optimisation-based energy system models, with examples including energy system analyses for Norway [6], Denmark [7], or the United Kingdom [8]. These previous applications have typically relied on relatively small scenario sample sizes (often below 500), which may be insufficient to obtain stable and representative output distributions for large-scale energy system optimisation models with extensive parameter spaces.

Against this background, we introduce a methodological workflow that couples temporally conditional Monte-Carlo simulation with a national energy system optimisation model to systematically assess uncertainty in long-term energy system transformation pathways. The proposed approach enables the identification of probable transition trajectories, the estimation of minimum infrastructure expansion requirements, and the quantification of likely technology deployment levels under uncertain techno-economic assumptions. We apply our approach to two national case studies representing the pathways to greenhouse gas-neutrality in Germany and Great Britain (see Supplementary Notes 4 and 5). The results illustrate how the proposed framework can enhance the interpretability of energy system modelling results and provide decision-makers with a more comprehensive understanding of uncertainty in scenario-based energy transition analyses.

# Results

## Probability space in national energy system transformations

To establish a benchmark for interpreting the probabilistic model results, a standard scenario was defined for each country. These reference pathways are constructed using the mean values of all parameter distributions and therefore represents a deterministic transformation pathway comparable to those reported in other energy system studies.

The resulting electricity mixes in the two standard scenarios (see Figure 1) reflect both historical developments and strategic differences in national energy policies. The earlier coal phase-out in Great Britain, combined with the continued use of nuclear power, leads to a more rapid decarbonisation of the electricity sector. In the British case, more than 90% of electricity generation is decarbonised by around 2030, whereas Germany reaches a comparable level only at a later stage of the transition. This difference arises primarily because coal-fired power plants remain in operation in Germany until approximately 2035, while the continued use of gas-fired generation also slows the pace of decarbonisation. At the same time, differences in renewable resource potential shape the long-term technology mix. The favourable wind conditions in Great Britain result in a larger share of wind power generation compared to Germany, even though wind remains the largest renewable technology in both countries. Conversely, higher solar irradiation – particularly in southern Germany – combined with an already substantial installed base of photovoltaic (PV) capacity leads to a more prominent role for solar photovoltaics in the German energy system.

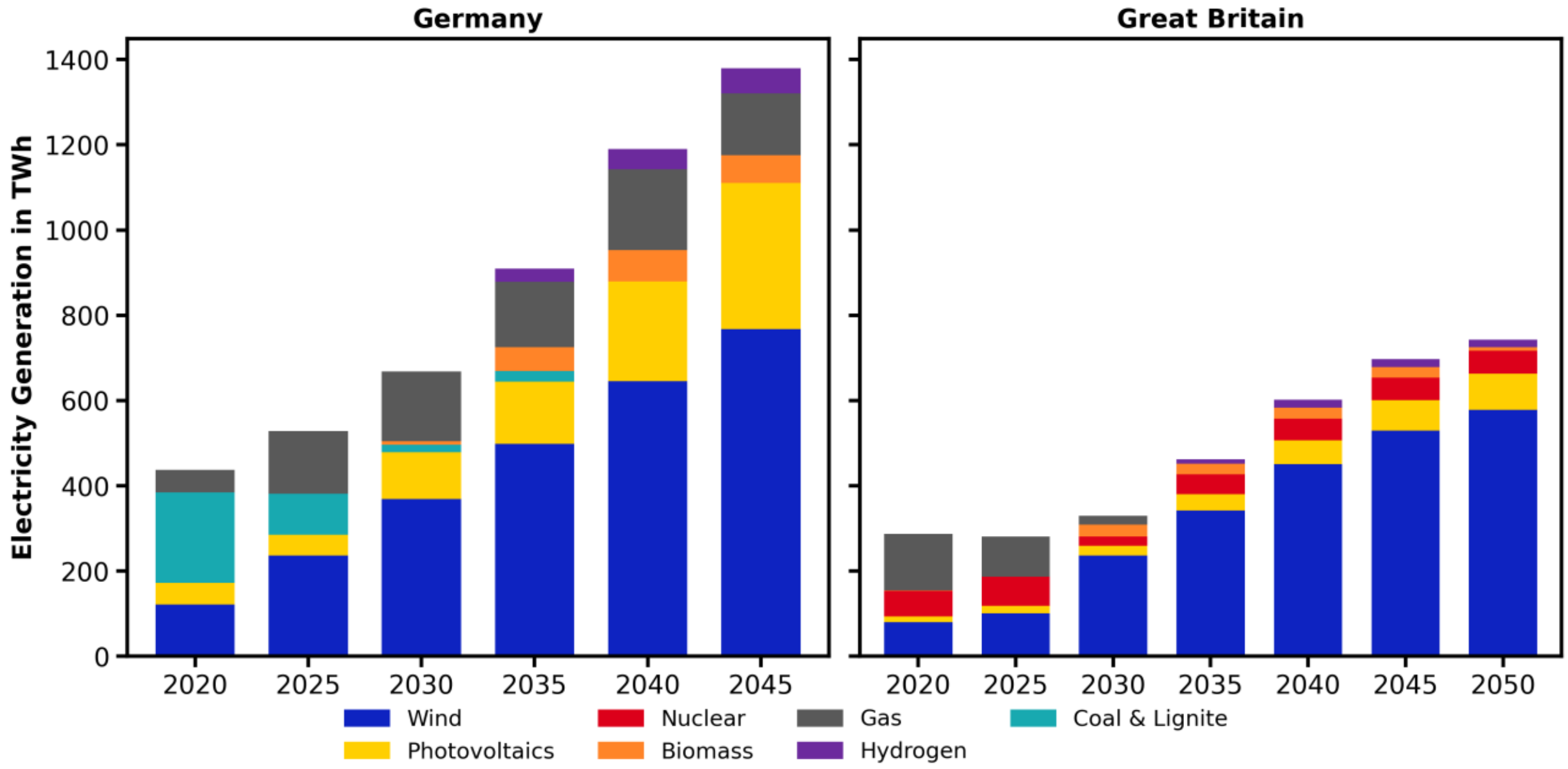

*Figure 1: Electricity generation in Germany between 2020 and 2045 (a) and Great Britain between 2020 and 2050 (b) divided by energy source for an optimised standard scenario using the mean values of the parameter distributions. The optimised standard scenario represents the optimal trajectory that follows the central cost and technology parameters, constrained to meet government targets für emissions reduction up to net-zero carbon emissions by 2045 in Germany and by 2050 in Great Britain. These results are model outputs of a cost optimisation model for the entire timeframe, hence why the historical electricity mix in the two countries in 2020 and 2025 might differ from the model outputs. This is particularly visible in Germany, where the model output did not incentivise using nuclear power in 2020.*

While the overall structural trends observed in the standard scenario remain visible across the ensemble of simulations, the probabilistic analysis with 10,000 individual scenarios for each country reveals substantial uncertainty in the future electricity mix (see Figure 2). Wind power becomes the central component of decarbonised electricity production in both countries. In the $5^{th}$ percentile of the distribution for Germany, wind generation still exceeds 400 TWh by 2045, indicating that scenarios with significantly lower wind deployment are highly unlikely under the assumption of greenhouse gas neutrality. Nevertheless, the difference of roughly 350 TWh between the $5^{th}$ and $95^{th}$ percentiles highlights the considerable uncertainty regarding the precise level of wind deployment required.

For Germany, solar photovoltaics power generation also exhibits very wide uncertainty between approximately 200-750 TWh in 2045. Although this spread indicates substantial flexibility in the optimal deployment of PV capacity, the lower bound of 200 TWh suggests that a German energy system without significant solar PV deployment is highly improbable. The majority of British scenarios (75%) show relatively limited PV utilization below 100 TWh. As a result, more than 70% of the scenarios show wind as the largest contributor to electricity supply in 2050. The comparatively lower capacity factors of PV in Great Britain, combined with investment costs that are insufficiently low to offset this disadvantage, limit the economic competitiveness of solar technologies.

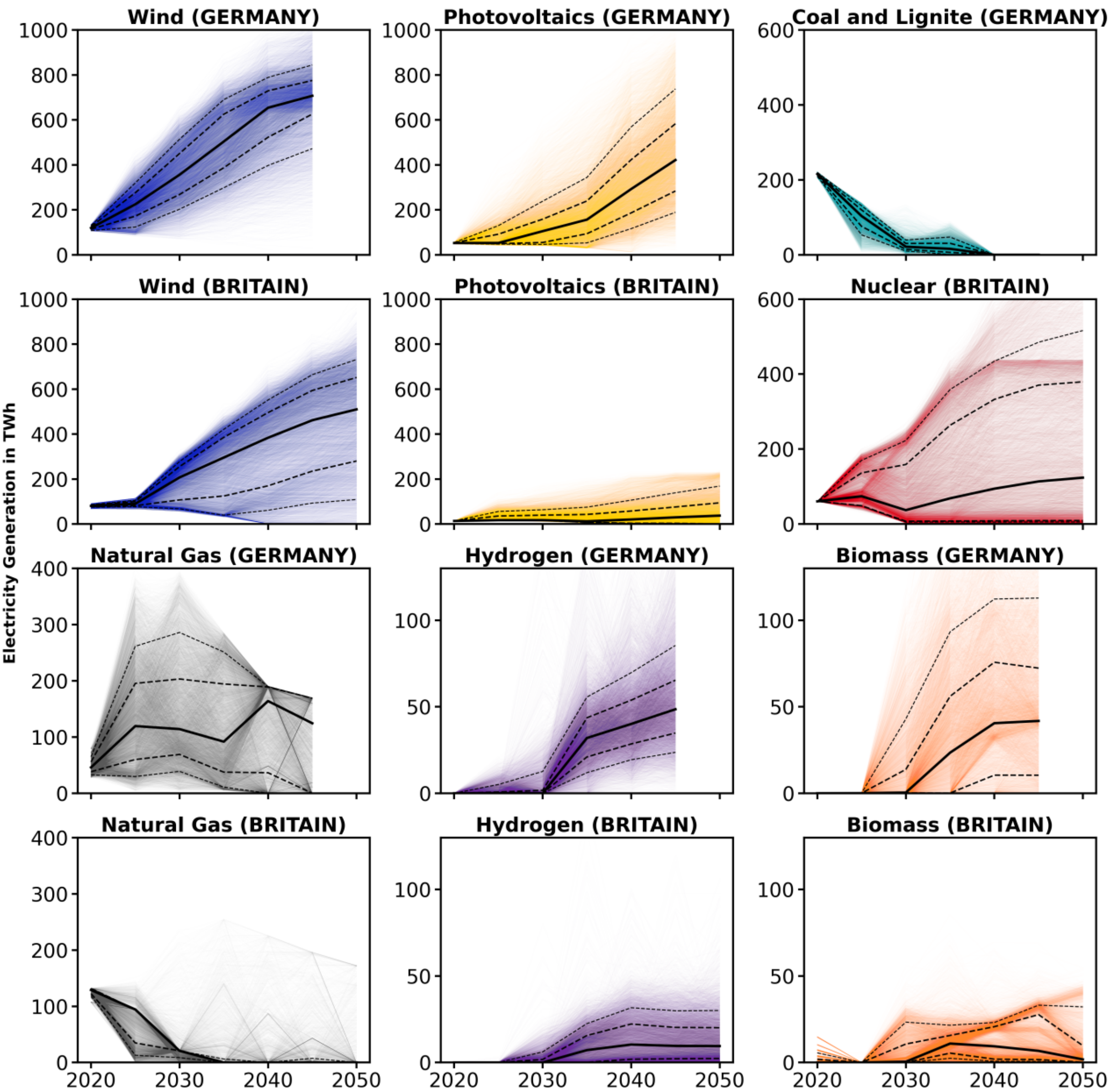

*Figure 2: Electricity generation in Germany and Great Britain over all 10 000 pathways split by energy source. The figure shows the increasing importance of renewable electricity generation contrasting the decline of fossil power usage as part of the energy transition. Dotted lines represent the 25$^{th}$ and 75$^{th}$ percentile as well as the 90$^{th}$ and 10$^{th}$ percentile, while the solid black line represents the median value. The system trajectories are modelled until 2045 for Germany and 2050 for Great Britain, consistent with the respective national targets for reaching net-zero greenhouse gas emissions.*

Dispatchable generation technologies – particularly gas-fired power plants – exhibit strong divergence between the two countries. In Germany, gas generation varies widely across scenarios, ranging from near-complete phase out to levels exceeding 200-300 TWh, driven by gas price variability, renewable costs, and the availability of carbon capture options. Gas with carbon capture and storage (CCS) plays a significant transitional role, especially around 2040, before declining toward 2045 under stricter emission limits. In Great Britain, by contrast, gas plays a more limited and shorter-lived role, with most scenarios showing a rapid phase-out by around 2035. The lower reliance on gas is largely due to the availability of alternative low-carbon baseload options, particularly nuclear power, and a lower attractiveness of gas with CCS. Biomass peaks in both systems occur under similar economic conditions (high nuclear costs, low gas prices, and higher costs for intermittent renewables), with a growing share using carbon capture from 2035 onwards. Overall, Great Britain transitions more quickly toward fully carbon-neutral baseload generation, whereas Germany maintains a more prolonged and uncertain role for dispatchable low-carbon technologies.

The potential role of nuclear power in Great Britain leads to one of the most pronounced structural differences between the two countries (Figure 2), with a large proportion of scenarios including significant nuclear capacity expansion. Two distinct pathways emerge within the scenario ensemble. One cluster of scenarios shows substantial nuclear expansion (up to around 400 TWh), while another cluster shows minimal or no deployment. This split reflects high sensitivity to nuclear investment costs, financing conditions, and competition from alternative technologies, and it significantly shapes the overall system configuration in Great Britain.

The phase-out of coal and lignite is highly robust in Germany, with narrow uncertainty bands reflecting the mandated phase-out by 2038. Only a small subset of scenarios shows a temporary resurgence around 2035, driven by a combination of high gas prices, high investment costs of renewable and gas-fired power plants, and low costs for carbon dioxide removal technologies such as bioenergy with carbon capture and storage (BECCS).

## Technology cost tipping points

The cost of each technology is the strongest driver of its uptake and output. Higher wind cost means less wind is built, for example. Most technologies follow a similar pattern: output is zero until costs fall below a certain threshold. Then, as costs fall further, their uptake increases linearly..

We calculated the threshold costs for all technologies in Britian and Germany. A walkthrough of the calculation is given in Figure 3, using the example of small modular nuclear reactors in Britain, and gas CCS in Germany. The model runs show that small modular reactors would be selected more often than not as early as 2030 if their capital cost was below €5010/kW. Some scenarios see them used at much higher costs (€6,600/kW at the 99$^{th}$ percentile), while others see no uptake even with much lower costs (€2,600/kW at the 1$^{st}$ percentile), showing the high interdependence on other system assumptions. Below this threshold cost, there is a moderately linear relationship with output increasing as capital cost decreases ($R^2$ = 0.40 in 2030, declining to 0.26 in 2050) as the slope becomes weaker (less variation in cost as a function of output in Figure 3b). The resulting threshold cost (Figure 3c) declines gradually from €5010±800/kW in 2030 to €4410±800/kW in 2040 and €4160±790/kW in 2050. The level of scatter in panels (a-c) indicates how strongly other factors affect the optimal output of a given technology, which in turn drives the level of uncertainty seen in the regressions and thus in the target cost (shaded area in panel c).

This provides a capex target for industry and RD&D programmes, which if met would likely yield an attractive technology for investors. The target for SMRs is towards the lower end of our (broad and uncertain) input cost estimates, and declines by ~1% per year, implying that continued cost reductions are required for continued uptake because of competition from other low-carbon technologies. In contrast, the target cost for gas CCS in Germany (Figure 3f) increase during the 2030s to meet the median of our input cost assumption. This reflects the German emission limits becoming more strict over time, which prevent unabated gas and thus increase demand for low-emission dispatchable capacity.

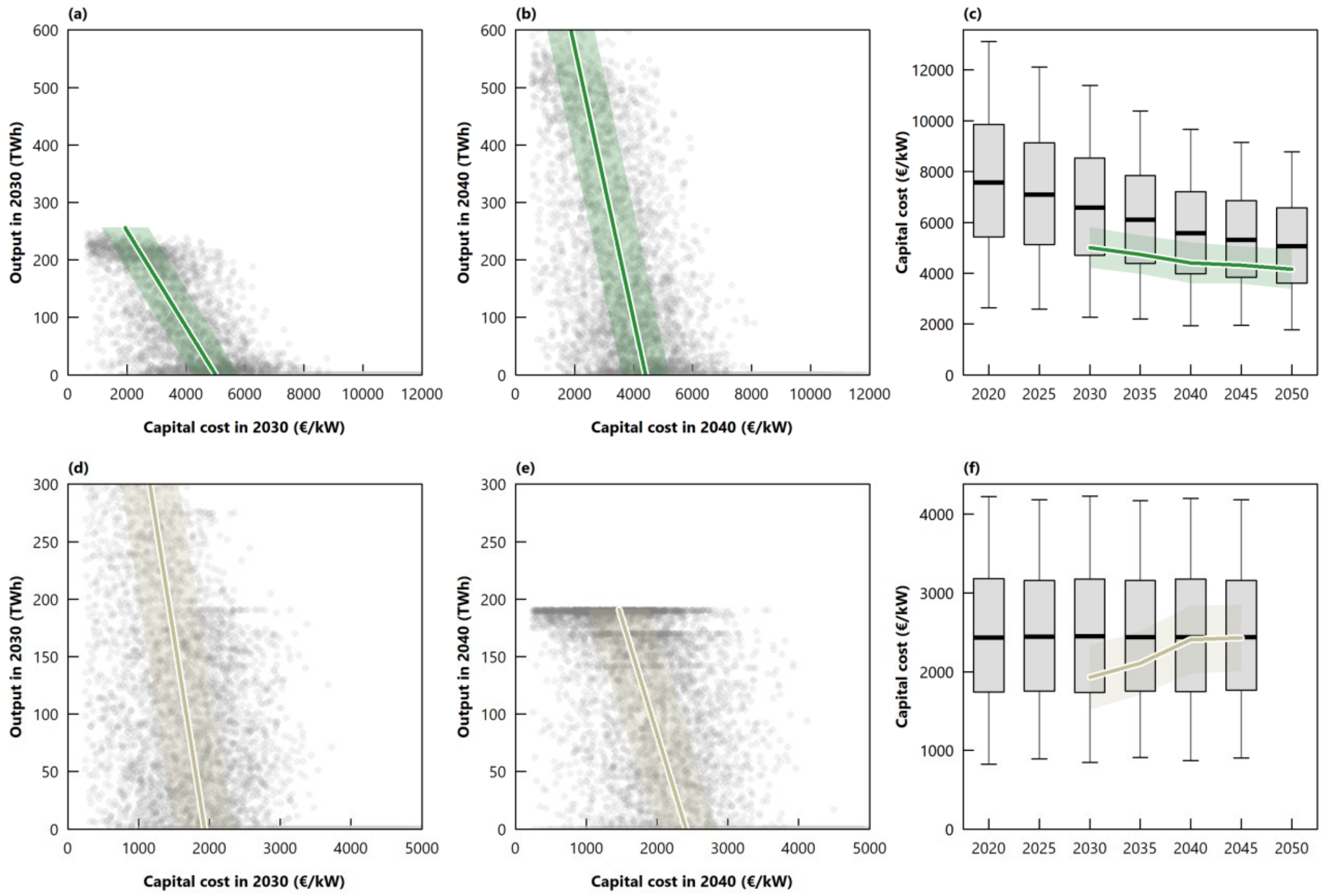

*Figure 3: Calculation of the cost tipping point for small modular nuclear reactors in Britain (top) and gas with CCS in Germany (bottom). Scatter plots show the relationship between capital cost and electricity output in 2030 (a, d) and 2040 (b, e). Points represent individual Monte Carlo trials, thick lines show the linear regression (which excludes zero values), and the shaded area represents the prediction interval of the least-squares regression. Boxplots show the range of capital cost inputs to the model, with line and area showing the corresponding threshold cost (c, f). Bars represent the interquartile range of cost inputs across all Monte Carlo trials, whiskers extend to the minimum and maximum, and thick black lines represent the median. Thick coloured lines represent the threshold cost below which technology uptake occurs, derived from the regressions of cost against output. Pale shaded areas represent the prediction interval. The corresponding plots for other technologies are given in Supplementary Figures 1–24.*

The approach from Figure 3 was applied to all technologies in Great Britain and Germany (see Figure 4). In both countries, electrolysers and solar PV have the lowest target costs. PEM electrolysis can be substituted by alkaline electrolysis, pyrolysis or hydrogen imports, hence its threshold cost must be low, and decline over time, in both countries to remain competitive. Solar PV has low threshold costs because of its low capacity factors relative to other generators. Its threshold cost declines over time in Britain due to increasing competition from wind or nuclear generation, while in Germany the threshold remains stable as solar (alongside onshore wind) outcompetes other technologies and remains in high demand.

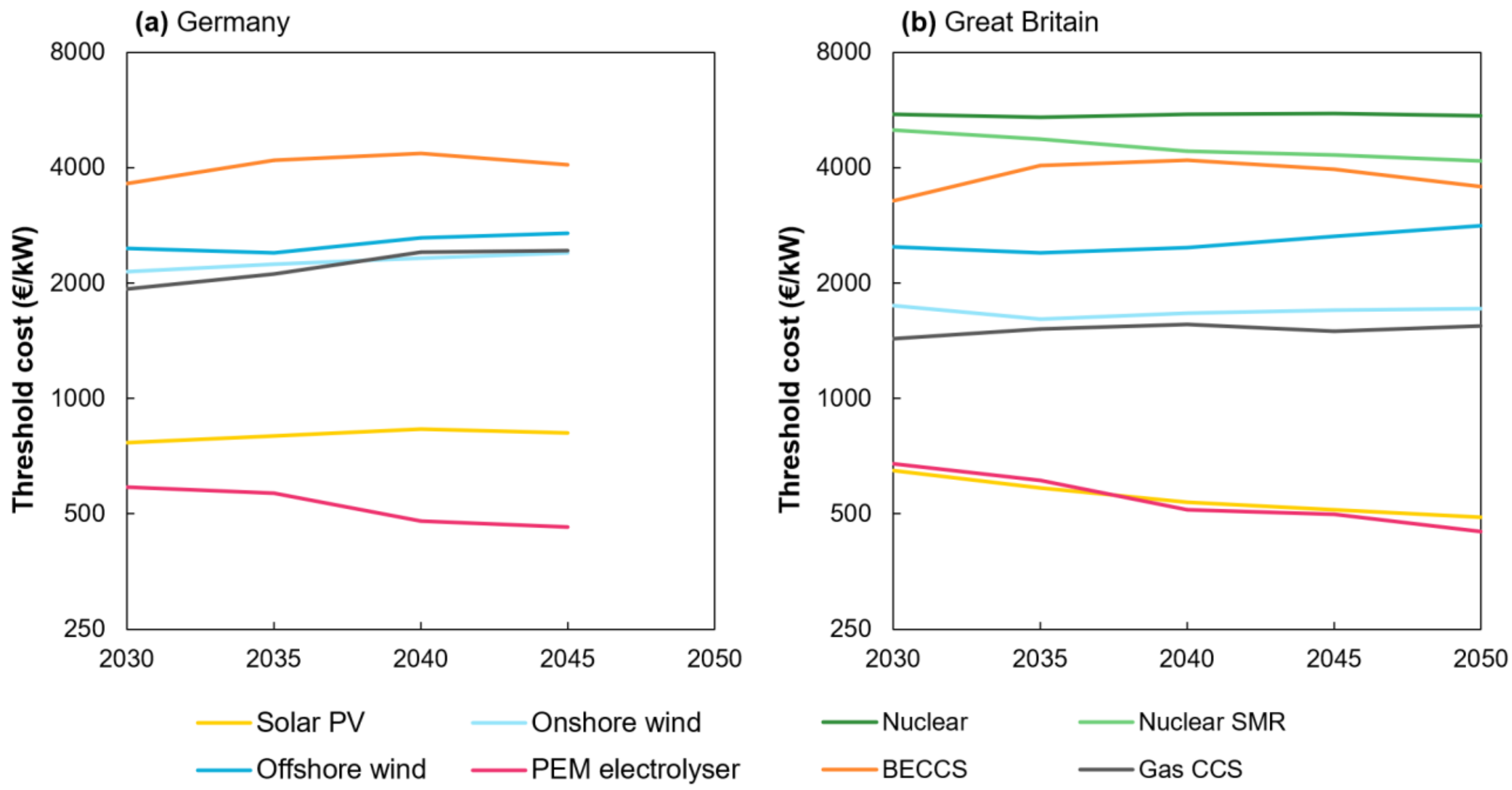

*Figure 4: Capital cost thresholds below which technologies are adopted in Germany and Great Britain as a function of time.*

Offshore wind has a 49% higher cost threshold than onshore wind in Britain, versus just a 12% premium in Germany. In Britain, this is largely explained by their relative productivity, with an average capacity factor of 45%, offshore wind can remain competitive at higher cost than onshore wind (averaging 32% capacity factor). Germany's offshore wind resources have comparatively low capacity factors (averaging 33%), and the total resource is limited to ~200 TWh/yr by available sea floor. Offshore wind output is quickly exhausted as its cost falls, which depresses the threshold cost relative to onshore wind, which is required to meet the bulk of electricity demand. The cost thresholds for wind in both countries also increase over time, indicating a greater willingness to pay to ensure the system to meet legally-binding carbon reduction targets, as wind power supplies the bulk of decarbonisation in both countries (offshore in Britain, onshore in Germany).

The cost thresholds for gas with CCS track at a similar level to onshore wind in both countries. Nuclear threshold costs in Britain are notably higher, although not as high as current construction costs. The threshold is lower for small modular reactors, and falls further over time, due to the efficiency and operating cost advantages of conventional reactors. In the absence of nuclear power, negative emissions technologies have the highest threshold cost in Germany. Capital costs around €4000/kW are sufficiently competitive to spur uptake, as they are required to offset residual emissions in other sectors. Initially, this cost rises as both countries' electricity sectors face increasing pressure to deliver negative emissions from 2035 onwards, in order to remain consistent with economy-wide targets (see projections in the Methods). Growing competition from direct air capture (DAC) begins to push down the threshold costs after 2040.

## The factors driving competition between technologies

Our results highlight that a limited number of key parameters can distinguish clusters of different scenario archetypes and exert a disproportionate influence on the overall energy system design, while many other uncertain parameters have comparatively minor effects. In both Germany and Britain, investment costs of technologies emerge as dominant determinants of system configurations (Figure 5). General correlation coefficients between a technology's investment cost and its electricity generation typically range from $r$ = –0.30 to –0.73, making them substantially stronger than correlations associated with most other input parameters. The strongest cost sensitivities are observed for renewable electricity technologies and nuclear power, highlighting the central role of capital costs in shaping the optimal electricity mix. The influence becomes even more pronounced along the transformation pathway, as progressively stricter emission constraints increase the structural dependence of energy systems on low-carbon electricity generation (see Supplementary Figures 25–27).

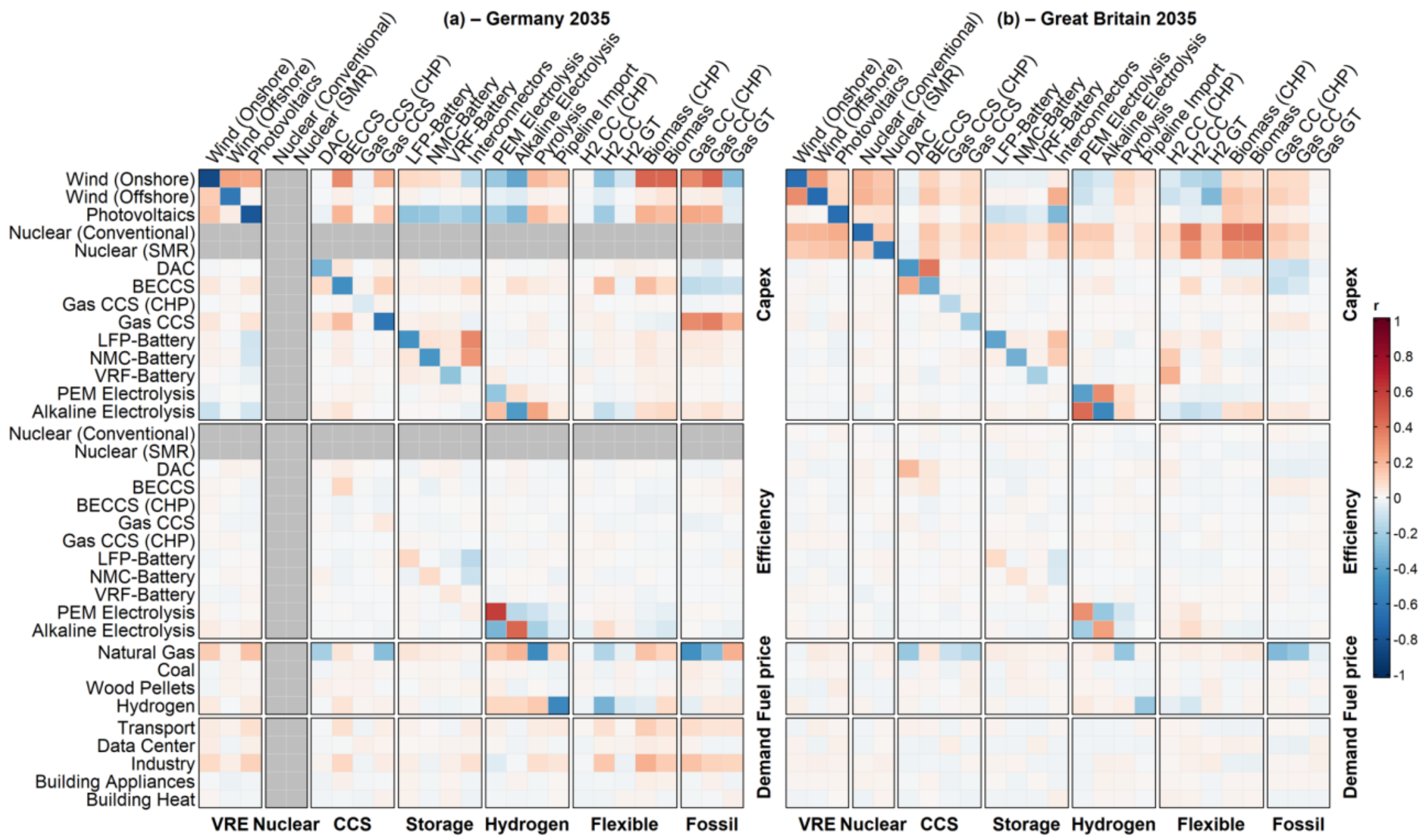

*Figure 5: Correlation matrices showing the impact of cost, efficiency and other input parameters on the output of technologies in 2035 for Germany (a) and Great Britain (b). The corresponding matrices for the years 2030 to 2045 are given in Supplementary Figures 25–27. Visualised using [9]. The rows of the matrices are displaying the model's input parameters grouped by capital expenditure (Capex) and efficiency, as well as fuel prices and electricity demands. The columns are displaying the model's output results grouped by characteristics of the technology. The*

***abbreviations in the figure are VRE: Variable Renewable Energy, CCS: Carbon Capture and Storage, SMR: Small Modular Reactor, BECCS: Biomass Energy with Carbon Capture and Storage, CHP: Combined Heat and Power, LFP: Lithium-Iron Phosphate, NMC: Nickel-Mangan-Cobalt, VRF: Vanadium Redox Flow, PEM, Polymer Electrolyte Membrane, CC: Combined Cycle, GT: Gas Turbine.***

In addition to a technology's own cost, the costs of directly competing technologies strongly influence deployment outcomes. This competitive interaction is particularly evident for onshore versus offshore wind power, as well as for conventional nuclear reactors versus small modular reactors. In these cases, higher costs for one technology are systematically associated with increased deployment of the other technology, reflecting direct substitution effects within the optimisation framework.

The competition between forms of variable renewable energy is strongly influenced by their relative capital costs (Figure 6). In 2035, the dividing line between offshore and onshore wind providing a quarter of electricity supply (vertical lines in Figure 6) occurs when they have comparable costs in Germany, and when offshore wind has a cost 1.50 times that of offshore wind in Britain, aligning with the difference in threshold costs from Figure 4. Similarly, the cost of solar PV relative to wind technologies identifies conditions in which it will provide a quarter of electricity: when it costs less than 27% of either wind technology in Germany; or in Great Britain when solar cost is less than 20% of onshore wind (horizontal lines), or 13% of offshore wind (diagonal lines).

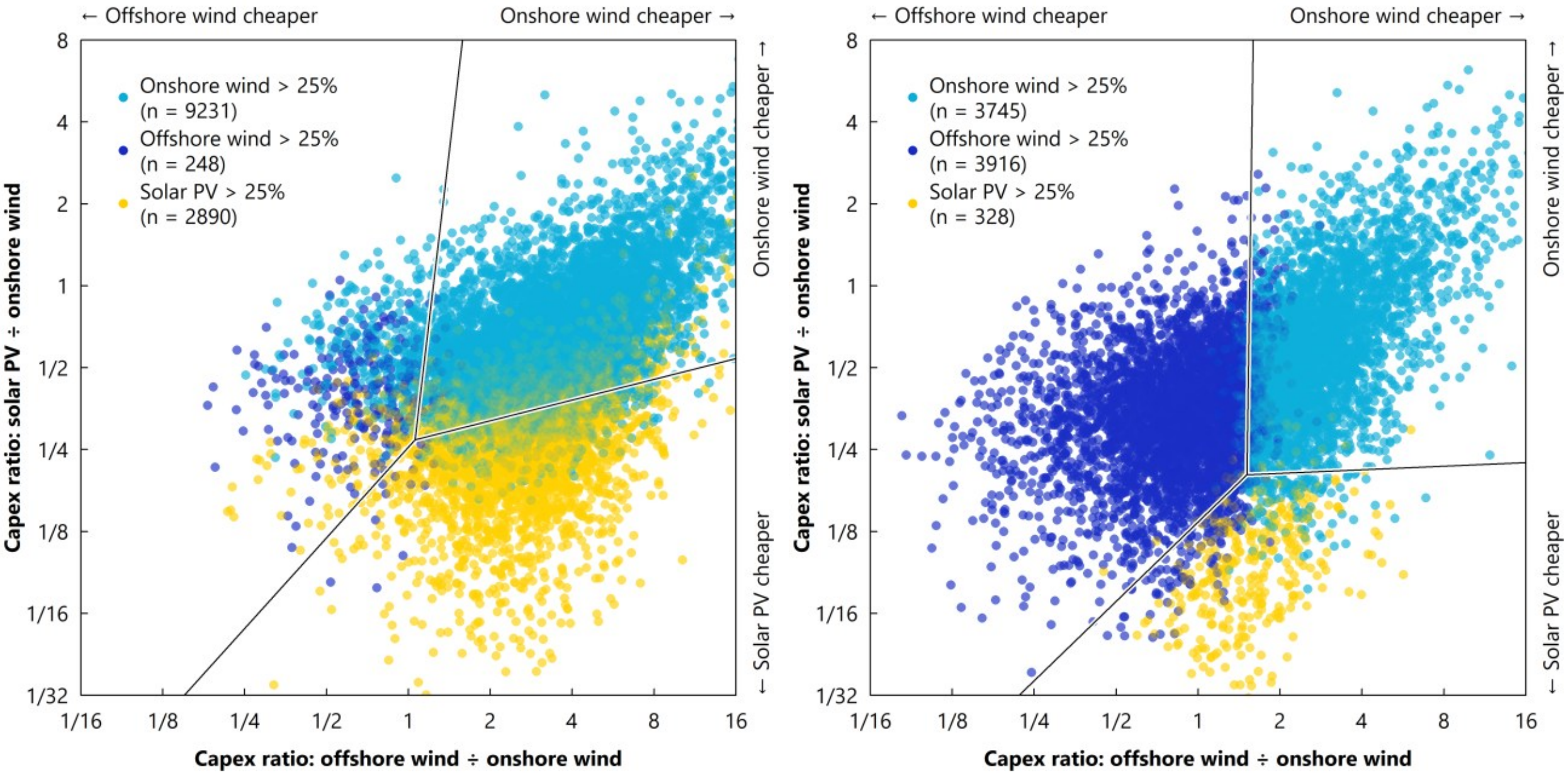

***Figure 6: Model runs with the highest shares of each form of renewable generation form distinct clusters based on their relative costs. Each point represents an individual model run for 2035 in Germany (left) and Great Britain (right), in which one of the three technologies shown supplies more than a quarter of electricity demand, coloured according to which technology. Black lines indicate the decision boundaries between individual technologies being dominant, identified via linear discriminant analysis (LDA).***

Fuel prices are also a key determinant of system outcomes, with the German energy system showing high sensitivity to natural gas price variations across multiple technologies. Higher gas prices increase the competitiveness of renewables. PV deployment is more strongly affected by gas prices compared to wind due to the higher intermittency of PV and the resulting need for flexibility – which could be provided by dispatchable gas power plants. Additionally, this also drives battery storage deployment based on the same need for flexibility, reflecting its role in stabilising PV-heavy generation profiles.

Natural gas prices also influence the deployment of carbon dioxide removal technologies, particularly direct air capture. Lower gas prices increase the profitability of gas-fired power generation, as indicated by the negative correlation between gas prices and gas-fired generation. Because gas-fired power plants – even when equipped with CCS – continue to emit residual $CO_2$, higher gas utilisation increases the need for carbon dioxide removal. As a result, scenarios with lower gas prices tend to exhibit higher DAC deployment. Additionally, in the model framework the price of natural gas is linked to the price of heat, which constitutes a key input for DAC processes. Lower gas prices therefore reduce the operational costs of DAC systems and further strengthen this relationship.

The hydrogen sector is also strongly influenced by natural gas prices. Hydrogen can be produced via electrolysis or through methane pyrolysis, the latter relying on natural gas as a feedstock. While hydrogen-fired power generation might be expected to benefit from higher gas prices due to competition with gas-fired plants, the availability of cheap hydrogen produced via methane pyrolysis under low gas price conditions ultimately dominates this effect. The price of imported hydrogen also influences hydrogen-related technologies, although its effect is generally weaker than that of natural gas prices. This effect is stronger in Germany in 2035 (Figure 6), but as Great Britain is projected to become a net hydrogen exporter, import prices exhibit somewhat stronger correlations with domestic electrolysis and pyrolysis deployment by 2045 (Supplementary Figure 27).

Some other factors, such as fuel prices of coal and biomass, technology efficiencies, or variations in electricity demand across different sectors, exhibit relatively limited influence on the overall energy system configuration. Data centre electricity demand, which has received considerable attention in recent policy discussions due to the rapid expansion of artificial intelligence infrastructure, exhibits only weak correlations with most system outputs as industrial electricity demand is significantly larger in absolute terms in both countries.

Although indirect substitution effects exist – for example between PV combined with battery storage and flexible thermal generation – the strongest competitive relationships occur between direct substitute technologies, as described above for variable renewable energy technologies. Similar substitution patterns appear across several technology pairs, including DAC versus BECCS, different lithium-ion battery chemistries (e.g. LFP versus NMC), and alkaline versus PEM electrolysers. In each case, positive off-diagonal correlations in the cost-deployment matrix indicate that higher costs for one technology increase the attractiveness of its direct substitute.

# Discussion

Deep decarbonisation pathways are increasingly shaped by technologies whose costs, performance, and scalability remain uncertain, yet standard practice in energy system optimisation still too often communicates a small number of deterministic, cost-optimal futures. We reframe pathway analysis from selecting a single "most likely" portfolio to characterising which features of the transition are robust necessities, which are contingent choices, and where the system exhibits tipping-point behaviour that is decision-relevant for near-term strategy.

Across German and British case studies which seek net-zero emissions by mid-century, the probabilistic pathways reveal a combination of structural regularities and divergent branching. Renewable expansion is consistently central to meeting tightening emissions budgets, but the composition and timing vary widely across the Monte Carlo ensemble, producing broad uncertainty envelopes for generation mixes even when the end goal is fixed. Britain exhibits a pronounced bifurcation in which nuclear expansion is a high-probability outcome in many runs while remaining negligible in others, depending primarily on the capital cost of reactors and other low-carbon technologies. In contrast, the German scenario ,where building new nuclear reactors is forbidden, shows compensating shifts toward other forms of firm or dispatchable supply when renewables become less favourable. Multiple internally coherent system archetypes can arise from plausible combinations of assumptions: with high solar, wind, fossil-CCS, hydrogen, or negative emissions all represented (see Supplementary Note 6).

A central value of the probabilistic framing is that it yields outputs that can be mapped directly onto innovation and policy questions. By regressing deployment outcomes against sampled capital costs, the analysis identifies technology cost tipping points: threshold costs below which investment begins, expressed with uncertainty bands that reflect the wider system context. The illustrative case of small modular reactors in Britain shows uptake becoming economical around a threshold on the order of €5,000/kW in 2030 and €4,400/kW in 2050. These values are broadly consistent with previous modelling and meta-studies, which have identified competitiveness thresholds around $5,000–8,000/kW [10–12]. There was substantial variability around the median threshold values in our analysis – including deployment at much higher costs and non-uptake at lower costs – reflecting the fact that any one technology's competitiveness is conditional on the wider system configuration.

More broadly, our study finds that threshold costs evolve over time in ways consistent with tightening constraints: some technologies require continued cost reductions to retain competitiveness, while others can justify higher willingness-to-pay as they become system-critical for meeting binding emissions limits. More important than the absolute cost of a technology is its cost relative to competing technologies. Faster innovation and cost reductions in one technology will therefore increase its share in a national power system while decreasing the share of its competitors: for example, solar PV has greater deployment when wind is expensive, and also when batteries are cheap. The price of natural gas propagates widely through the system, affecting gas generation directly, and also indirectly the attractiveness of solar-plus-storage configurations, hydrogen production routes, and the scale of carbon dioxide removals required to offset residual emissions. This knowledge of feasible low-cost system and technology pathways is particularly relevant given heightened natural gas–related energy security risks [13].

Looking beyond costs, this work finds that variation in electricity demand (including from data centres) has limited effect on system design compared to capital and fuel cost uncertainties. This provides a useful counter to narratives that treat load-growth uncertainty as the dominant planning challenge, at least within the demand magnitudes and representations applied here. While costs are a major driver of technology adoption in our macroeconomic analysis, efficiency can become equally important in space-constrained settings, such as for homeowners with limited roof space for solar PV [14]. Taken together, our results point to a small set of high-impact technology parameters that disproportionately shape portfolio outcomes and therefore merit priority in data collection, policy design, and risk management.

Future research can build on this foundation in three complementary directions. First, extending probabilistic pathway mapping to spatially resolved or multi-node models would test how transmission bottlenecks and regional resource quality affect tipping points and their resulting portfolios. Second, incorporating richer uncertainty structure, including correlated cost drivers and learning dynamics, would better capture the joint risks that policymakers face (for example, simultaneous shortfalls in multiple technologies). Third, there is a clear opportunity to connect these probabilistic outputs to decision frameworks: designing policies that deliberately move probability mass away from high-risk archetypes (e.g. portfolios reliant on narrow sets of assumptions) and toward strategies that are robust across the ensemble, while prioritising "value of information" efforts where uncertainty reduction would most improve decisions.

The broader conclusion from this work is that the relevant question for deeply decarbonised systems is no longer which single pathway is optimal under assumed futures, but which near-term actions most reliably keep the transition within a feasible envelope as the future unfolds. The framework we demonstrate with wide uncertainty bounds is an approach to assessing such actions across a wide envelope of possibilities.

# Methodology

## The ETHOS.NESTOR energy system model

A brownfield energy system optimisation model is used for assessing the uncertainty of the transformation pathways of the energy systems in Germany and Great Britain. The national energy system optimisation model ETHOS.NESTOR [15,16], was adapted to the research questions and the country-specific differences. We created two novel model instances, hereafter called *NESTOR-DE* and *NESTOR-GB* which are temporally resolved capacity expansion models able to optimise the capacity expansion and operation of energy technologies for multi-year periods (see Figure 7). The *NESTOR-DE* and *NESTOR-GB* models typically required between 10-25 minutes per model run on a high-performance computer (32 × Intel Xeon Gold 6334 @ 3.6GHz CPUs and 4 TB RAM).

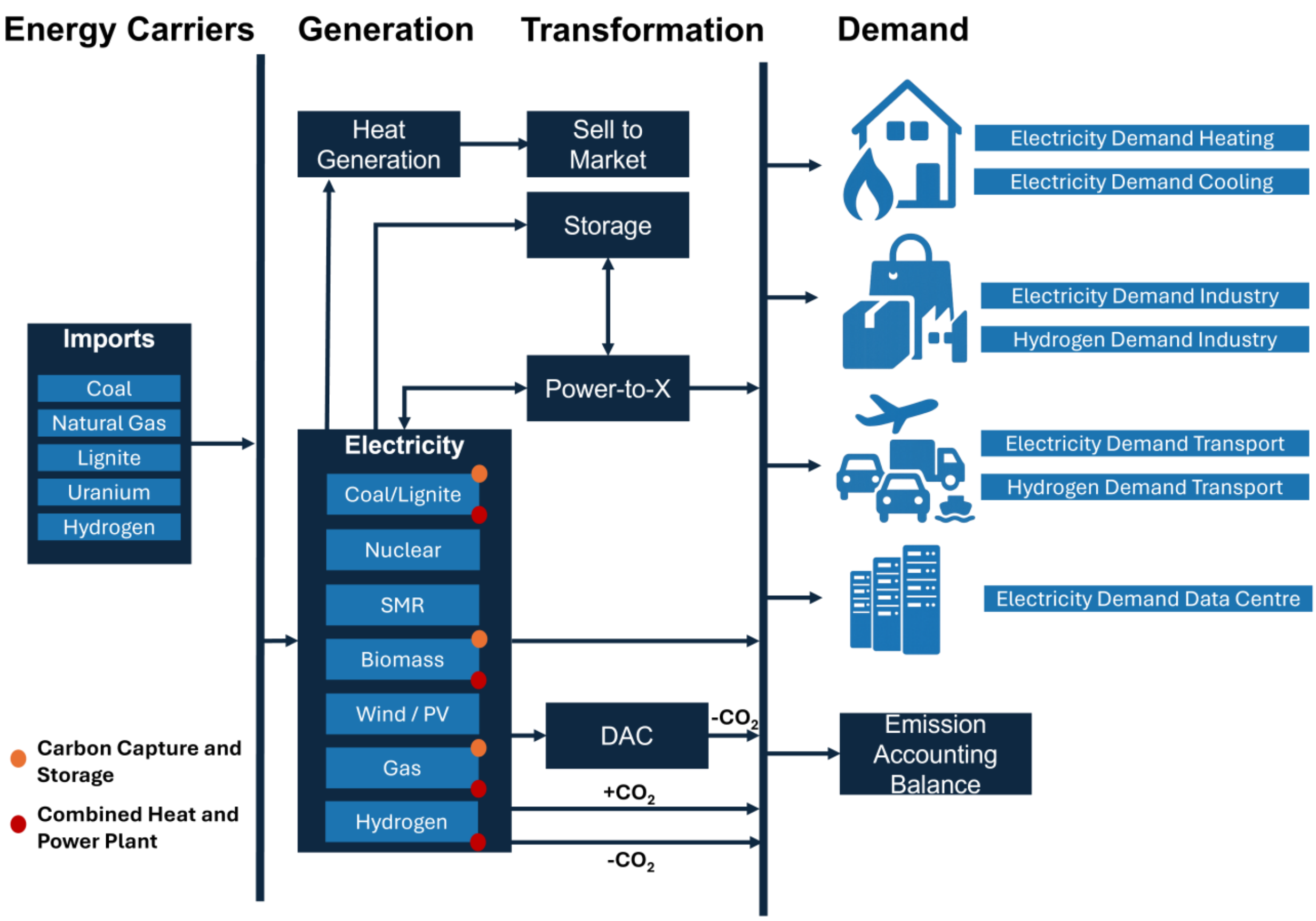

*Figure 7: Overview of the general concept of ETHOS.NESTOR. Electricity sources are split by different technologies. Red and orange dots next to the technology indicate that besides normal power plants, the model includes also the technology with either CCS or CHP. Heat as a by-product from these power plants can be sold at an average price of how much it would have cost to generate this heat otherwise.*

The application of Monte-Carlo based methods to large-scale optimisation models is inherently associated with substantial computational requirements [17]. Consequently, additional modelling strategies are necessary to ensure that extensive probabilistic analyses remain computationally tractable. To address this challenge, the national model variants *NESTOR-DE* and *NESTOR-GB* apply a sectoral aggregation approach, where the endogenous scope of both model variants includes the electricity generation system and closely coupled technologies, including hydrogen production and negative emission technologies. Demand-side sectors are, by contrast, incorporated through aggregated demand profiles and sector emission constraints, thereby capturing their system-wide impacts while maintaining a manageable computational burden.. The comparison with the higher-resolution *ETHOS.NESTOR* model reveals highly similar power system transformation pathways in both models.

To represent the existing electricity system, a brownfield modelling approach is adopted, ensuring that all relevant installed capacities are explicitly considered. For *NESTOR-DE*, this includes lignite- and coal-fired power plants together with their respective decommissioning schedules, as well as nuclear, biomass, natural gas power plants and the currently installed renewables. In contrast, *NESTOR-GB* reflects the completed coal phase-out in Great Britain and therefore excludes coal-based generation technologies. However, unlike in the German case, the construction of new nuclear reactors is permitted in the British model. Power plants within both models can procure primary energy carriers – including coal, lignite, and natural gas – at market prices.

To manage variability in renewable electricity generation and to enable sector coupling, the models include several electricity storage and conversion technologies. Short- and medium-term storage is represented by lithium-ion batteries and vanadium redox flow batteries, while water electrolysers provide a pathway for converting surplus electricity into hydrogen for chemical energy storage or further use in industry or transport. Both alkaline and proton exchange membrane electrolysers are included. The resulting hydrogen can be stored either in small-scale compressed storage tanks or in large-scale underground salt caverns, enabling both short-term balancing and long-term energy storage.

The deployment of renewable electricity generation technologies is constrained by country-specific technical potential limits derived from spatial analyses that incorporate multiple exclusion criteria. For Germany, renewable capacity potentials are based on the assessment by Risch et al. [18], while for Great Britain they are derived from Barrett et al. for solar [19] and Bosch et al. for wind [20,21]. The expansion of electrolysis capacity is further constrained by expected future manufacturing capacities in Germany and Great Britain, as well as by potential material bottlenecks, most notably the availability of iridium required for PEM electrolysers [22–24].

A key feature of using the *ETHOS.FINE* framework to build *NESTOR-DE* and *NESTOR-GB* is its integration with *TSAM* [25], which enables the reduction of temporal resolution while preserving representative system dynamics. In this study, a time series aggregation of 25 representative days per year with 12 segments per day is employed, based on the trade-off between accuracy and computation time. This reduces the original 8,760 hourly time steps to 300 representative ones, which are mapped by *TSAM* to approximate the full temporal variability of the energy system.

In addition to optimising intra-annual system operation, the models function as capacity expansion models with multi-period investment optimisation. The modelling horizon begins in 2020 and proceeds in five-year intervals. For Germany, the optimisation extends to 2045 while for Great Britain it extends to 2050, corresponding to the target year for greenhouse gas neutrality in each country [26,27].
To capture the variability of renewable generation and weather-dependent demand profiles, country-specific meteorological datasets are employed. These profiles are generated using *Renewables.ninja*, based on 45 years of historical weather data spanning 1980-2024, thereby ensuring a robust representation of interannual weather variability in the modelling framework [28,29].

## Monte Carlo approach

The objective of the Monte-Carlo simulation in this study is to generate consistent parameter sets that serve as inputs to the energy system optimisation model. For each parameter set, all model parameters associated with substantial uncertainty are sampled from their respective probability distributions. These distributions are derived from an extensive literature research, which provides mean values and plausible parameter ranges for all relevant technologies and model years.

To represent parameter uncertainty, Gaussian probability distributions are employed. The central value obtained from the literature review serves as the expectation value of the distribution, while the parameter ranges are interpreted as the 5$^{th}$ and 95$^{th}$ percentiles, thereby defining the spread of the distribution. For parameters that are not central to the research questions of this study, these literature-based best estimates are used directly. In contrast, deliberately wider uncertainty ranges are applied to parameters of particular analytical interest to capture a broader spectrum of plausible technological developments.

In total, 10,000 model runs are performed for both *NESTOR-DE* and *NESTOR-GB*, with variations applied to techno-economic parameters such as technology costs, efficiencies, energy demand level, weighted average cost of capital, and the weather year. Depending on the nature of the parameter, uncertainty ranges are specified either in absolute terms or relative deviations from the mean value. Absolute ranges are primarily used for parameters such as future energy demand, where the uncertainty range reflects the spread across projections reported in the literature. Relative ranges, by contrast, are applied to parameters such as technology costs or efficiencies, where uncertainty is more meaningfully expressed as a proportional deviation from the central estimate. This approach enables the application of consistent uncertainty structures across both national models, even when the underlying expected parameter values differ between Germany and Great Britain.

The uncertainty ranges for electricity generation and storage technologies are derived from the *GNESTE* database [30], while techno-economic assumptions for water electrolysers are based on aggregated findings from multiple meta-studies [31–34]. For direct air capture technologies, uncertainty ranges are adopted from Wenzel et al. [35]. Future electricity demand projections and their associated uncertainty for Great Britain are derived from the four future energy scenarios presented by NESO [36], with the resulting range interpreted as the $5^{th}$ and $95^{th}$ percentiles. For Germany, demand uncertainty is estimated from the standard deviation across projections reported in different German energy system studies [37–43].

A key methodological innovation of the sampling approach implemented in this study is the introduction of temporal conditionality in the parameter generation process. Temporal conditionality is essential when analysing transformation pathways in long-term energy system optimisation models, as it ensures that sampled parameter trajectories remain technologically plausible over time. For example, if the sampling procedure yields an electrolyser efficiency of $65\%_{LHV}$ for PEM electrolysers in 2030, subsequent model years must not exhibit lower efficiencies. Enforcing such temporal consistency substantially reduces the feasible parameter space, thereby improving the realism of the sampled technology trajectories and reducing the computation time for sampling.

A straightforward method for enforcing such constraints is rejection sampling, in which a sampled value for year $i$ is compared with the previously drawn value for year $i\text{-}1$. If the newly sampled value violates the temporal constraint, it is rejected and a new value is drawn until the condition is satisfied. However, this approach presents two major drawbacks. First, rejection sampling can become computationally inefficient, particularly when feasible values are rare relative to the underlying distribution. Second, repeated rejection can introduce distributional distortions, potentially leading to artificial lock-in effects in which the resulting distributions become skewed toward the constraint boundary.

To address these limitations, this study employs Gaussian copulas to impose temporal dependence between parameters across model years while preserving the shape of the marginal distributions for every year. The approach begins by defining truncated Gaussian marginal distributions for each parameter and year, where the probability density function is rescaled so that values are only defined for $x \geq x_{min}$.

$$f_{trunc}(x) = \frac{f(x)}{\int_{x_{min}}^{\infty} f(t)dt}$$

Each truncated marginal distribution is then mapped to a uniform distribution on the interval [0,1] using its cumulative density function (CDF). In the uniform space the correlation between the marginals is imposed by sorting the uniform values in descending order. Sorting in uniform space preserves the distribution shape when mapped back. Then, the inverse CDF is used to map the uniform values back to the original distribution. The trade-off can be managed by introducing a monotonicity factor $\rho$ into the equicorrelation matrix $\Sigma$ which defines the correlation between the marginals in the Gaussian copula. A value of $\rho = 1$ would imply perfect correlation, guaranteeing strictly monotonic trajectories across time. However, such a setting can lead to numerical instability and unrealistic rigidity in the sampled parameter paths. In practice, a monotonicity factor of $\rho = 0.9$ was found to provide an effective compromise, preserving strong temporal consistency while maintaining the statistical properties of the marginal distributions.

For parameters that do not exhibit a strictly monotonic temporal evolution – such as fuel prices – no temporal dependency structure is imposed. Instead, these parameters are sampled independently for each model year from their respective marginal distributions.

## Demand and emissions projections

As we focus primarily on emissions from the electricity sector, we take the emissions budget available to that sector as an input, rather than the economy-wide targets of net-zero by 2050 or 2045 (Figure 8). For Britain, we use the sum of $CO_2$ emissions from electricity supply and engineered removals emissions from the CCC 7$^{th}$ Carbon Budget [44], which recommends pathways to Britain government to meet its legally-binding emissions reduction target. These UK-wide values were reduced by 2.5% [45] to account for Northern Ireland being a separate power system, not included in this model. For Germany, we take the expected emissions pathway by summing the same sectors across all scenarios in [37–43], and taking the mean. In both countries, the emissions from 2020 were taken as the average of historical emissions in 2019 and 2021 [46,47], to remove the effect of the COVID19 pandemic.

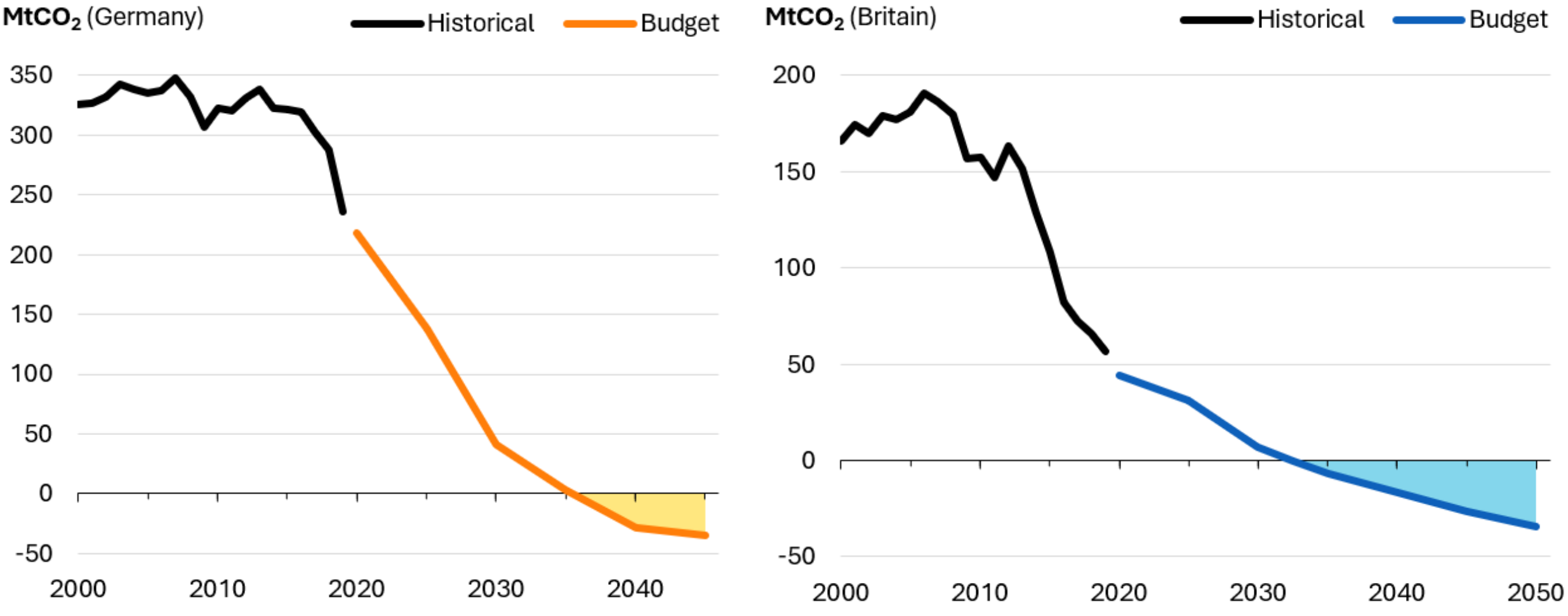

*Figure 8: National carbon emissions from electricity generation and removals in Germany (left) and Britain (right). Historical emissions are shown from 2000 [46,48], with the budget the model must adhere to. Net negative emissions are required beyond 2035 (Germany) and 2033 (Britain), highlighted with pale shading.*

The evolution of annual electricity demand was an exogenous input, specified for each year from 2020 through to 2050. Demand was specified separately for the five end uses shown in Figure 9, each with associated uncertainty ranges. For Britain, these were taken from NESO's Future Energy Scenarios [36], taking the mean and standard deviation across the three scenarios which align with the net zero targets (above). For Germany, the mean and standard deviation were taken across a set of representative energy system studies from different publications [37–43].

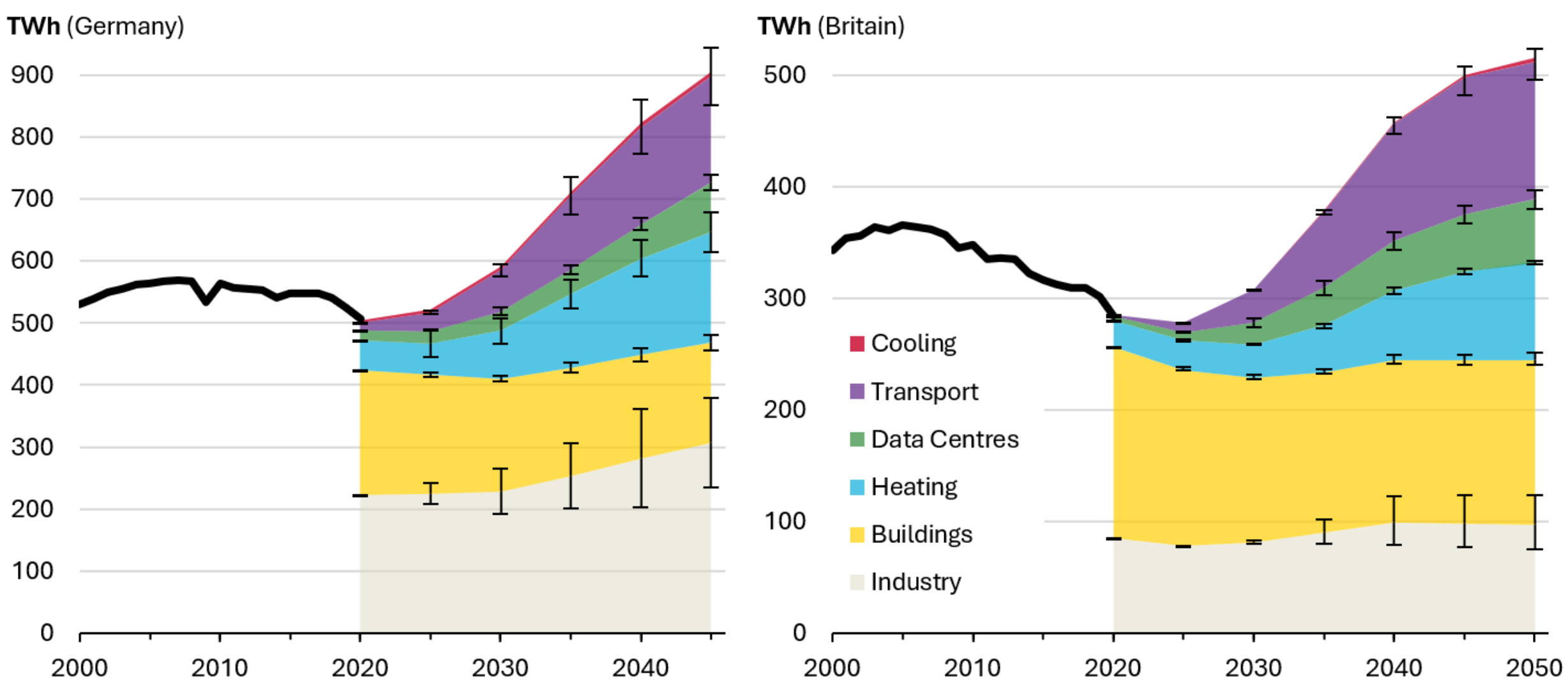

*Figure 9: Annual electricity demand in Germany (left) and Britain (right). Historical total demand is shown since 2000 [46,48], with projected demand disaggregated by the different end-uses implemented in the model. Uncertainty bars show the standard deviation on each demand component.*

## Time series data for weather-driven supply and demand

The time series of electricity production from solar PV, onshore and offshore wind were specified exogenously. These were modelled using the Renewables.ninja platform [28,29], using historical hourly irradiance, wind speed and temperature data from 1980 through to 2024. The national fleets of solar farms (both rooftop and utility-scale ground mounted) were simulated in each country, yielding annual average capacity factors of 12.5±0.4% in Germany and 10.9±0.3% in Britain. The projected future fleets of wind farms were simulated, based on existing farms plus the commercial pipeline of projects under construction or with approval as of January 2025, following the process in [49]. Onshore and offshore wind were simulated separately, yielding annual average capacity factors of 21.3±1.4% in Germany and 31.7±2.2% in Great Britain for onshore wind, and 32.6±1.9% in Germany and 45.0±2.2% in Britain for offshore wind.

Electricity demand from each sector was modelled as exogenous. For industry and data centres, demand was assumed to be constant across all hours of the year. For electric vehicles, the hourly profile from [50] was used for both countries, as it is represents metered demand from 8 million real-world charging events. For the residential and commercial sectors, hourly electricity demand from buildings, including their heating and cooling, was modelled using *Demand.ninja* [51]. This converts temperature, irradiance, wind speed and humidity data from 1980 to 2024 into modified heating and cooling degree-days. These are calibrated against measured historical national electricity demand in both countries to capture the pattern of consumption (incorporating weekends and national holidays). These patterns were then scaled up according to future annual total demand components.

## Technology cost projections

We take the investment costs for renewables and other power plants from the *EPD2030* study of the German energy transition [52] and the *SEDOS* database [53] (Table 1). Current cost levels were taken from the *SEDOS* database, while the cost development over time was adapted from the *EPD2030* study, inflating all investment costs into €$_{2024}$. For solar photovoltaics, investment costs are a weighted average of rooftop and open-field installations. Country-specific investment costs are assumed for wind and solar between Germany and Great Britain. Hydrogen technology and storage costs were adapted from the *SEDOS* database. Values for direct air capture were taken from Wenzel et al. [54].

***Table 1: Technology-specific investment costs used in NESTOR-DE and NESTOR-GB. Cost assumptions are provided for the years 2020, 2030, 2040, and 2050, while values for intermediate years are determined by linear interpolation. Wind and solar have country-specific investment costs, whereas other technologies use the same values in Germany and Great Britain.***

| | 2020 | 2030 | 2040 | 2050 | Unit |
|---|---|---|---|---|---|
| Electrolysis (AEL) | 750 | 650 | 580 | 500 | €/kW |
| Electrolysis (PEM) | 925 | 650 | 500 | 400 | €/kW |
| Methane Pyrolysis | - | 900 | 800 | 700 | €/kW |
| Hydrogen Gas Turbine | - | 1021 | 925 | 833 | €/kW |
| Hydrogen Combined-Cycle | - | 1144 | 1104 | 1063 | €/kW |
| Hydrogen Combined Cycle CHP | - | 1386 | 1334 | 1281 | €/kW |
| Direct Air Capture | 1296 | 1047 | 882 | 550 | €/(t/a) |
| Biomass Steam Turbine | 3275 | 2963 | 2688 | 2438 | €/kW |
| Biomass Steam Turbine (with CCS) | 4575 | 4263 | 3988 | 3738 | €/kW |
| Biomass Steam Turbine CHP | 4250 | 4000 | 3875 | 3750 | €/kW |
| Biomass Steam Turbine CHP (with CCS) | 5550 | 5300 | 5175 | 5050 | €/kW |
| Gas Power Plant Gas Turbine | 506 | 506 | 506 | 506 | €/kW |
| Gas Power Plant Combined-Cycle | 1063 | 1063 | 1063 | 1063 | €/kW |
| Gas Power Plant Combined Cycle (with CCS) | 2431 | 2431 | 2431 | 2431 | €/kW |
| Gas Power Plant Combined-Cycle CHP | 1469 | 1469 | 1469 | 1469 | €/kW |
| Gas Power Plant Combined-Cycle CHP (with CCS) | 2838 | 2810 | 2784 | 2758 | €/kW |
| Coal Power Plant Steam Turbine | 1625 | 1688 | 1750 | 1813 | €/kW |
| Coal Power Plant Steam Turbine CHP | 2781 | 2781 | 2781 | 2781 | €/kW |
| Coal Power Plant Steam Turbine (with CCS) | 3375 | 3188 | 3188 | 3188 | €/kW |
| Lignite Power Plant Steam Turbine | 1875 | 1959 | 2041 | 2125 | €/kW |
| Lignite Power Plant Steam Turbine CHP | 2781 | 2514 | 2406 | 2308 | €/kW |

| | | | | | | |
|---|---|---|---|---|---|---|
| Nuclear Power Plant | | 5734 | 5585 | 5585 | 5585 | €/kW |
| Small Modular Reactor | | 7500 | 6500 | 5500 | 5000 | €/kW |
| Wind Onshore | GB: | 1430 | 1430 | 1430 | 1430 | €/kW |
| | DE: | 1401 | 1228 | 1141 | 1086 | €/kW |
| Wind Offshore | GB: | 2462 | 2105 | 2047 | 2047 | €/kW |
| | DE: | 4260 | 3757 | 3252 | 2749 | €/kW |
| Solar Photovoltaics | GB: | 853 | 612 | 476 | 397 | €/kW |
| | DE: | 927 | 665 | 517 | 431 | €/kW |

Fuel costs were derived from [52,55] are summarised in Table 2:

***Table 2: Variable costs of buying different types of fuel for the electricity generation technologies. Cost assumptions are provided for the years 2020, 2030, 2040, and 2050, while values for intermediate years are determined by linear interpolation.***

| | **2020** | **2030** | **2040** | **2050** | **Unit** |
|---|---|---|---|---|---|
| Wood Pellets | 50 | 50 | 50 | 50 | €/MWh |
| Natural Gas | 116 | 28.6 | 28.6 | 28.6 | €/MWh |
| Coal | 8.2 | 11.2 | 11.2 | 11.2 | €/MWh |
| Lignite | 9 | 8 | 8 | 8 | €/MWh |
| Hydrogen (Pipeline) | - | 148 | 123 | 97 | €/MWh |
| Hydrogen ($LH_2$) | - | 192 | 159 | 126 | €/MWh |
| Uranium | 2.1 | 2.1 | 2.1 | 2.1 | €/MWh |

For technologies of interest (the novel technologies described in Supplementary Note 2), a deliberately wide uncertainty range was chosen to reflect the novelty and allow investigation of cost tipping points.

## Data analysis methods

We define a technology cost tipping point as the cost below which the respective technology starts to become economically viable and thus is invested into by the optimisation. By having a large number of Monte Carlo runs which span a wide range of cost assumptions, we are able to identify these tipping points for all technologies. These costs reflect the full range of uncertainties across other input assumptions (technology costs, fuel costs, demand).

To calculate the threshold below which investment begins in a technology, we extract data on the output $Q_{t,y}$, from that technology (t) in a given year (y), and the technology's capital cost in that year, $K_{t,y}$. We observe that output increases linearly as the technology cost falls below a threshold value. This threshold is a range rather than a point value because of the uncertainty in all other parameters. We identify this cost threshold (or tipping point), $C_{t,y}$, through linear regression. First, we remove model runs where the output from newly-built capacity is zero, to exclude pre-existing wind and solar farms built before 2020. We then perform an ordinary least squares regression of capital cost as a function of output, with the cost tipping point being the intercept:

$$K_{t,y,i} = C_{t,y} + \beta_1 \, Q_{t,y,i} + \varepsilon_i, \qquad i \in \{ i : Q_{t,y,i} > 0 \}$$

Where: $\beta_1$ is the regression coefficient, ε is the error term, and i is the index of Monte Carlo runs [1 to 10000].

To identify the most influential input parameters, we quantify the statistical relationships between the uncertain model inputs and selected model outputs. Specifically, we compute correlations between each parameter assigned an uncertainty range and a set of output variables of interest. For instance, correlations are calculated between parameters such as the investment costs of solar photovoltaics and system outcomes such as the installed renewable generation capacity.

The correlations are evaluated using the Pearson correlation coefficient rather than the Spearman correlation. The Pearson metric is particularly suitable in this context because it captures linear relationships with directional interpretation, allowing both positive and negative correlations to be identified. This is important for interpreting inverse relationships, for example, the expected negative correlation between technology investment costs and installed capacity, where higher costs tend to reduce optimal deployment levels.

To provide a comprehensive overview of system behaviour, a set of key output indicators was defined, representing central characteristics of the evolving energy system (e.g., technology deployment levels, system costs, or efficiencies). Based on these indicators, a correlation matrix was constructed that includes the correlations between all uncertain input parameters and each output dimension individually. This approach enables a systematic exploration of parameter influence across the entire system, facilitating the identification of dominant drivers, structural relationships, and emerging system patterns. The temporal dynamics of the energy transition are accounted for in the correlation analysis by performing a separate analysis for each model year. This allows the influence of individual parameters to be tracked over time and reveals how parameter importance evolves throughout the transformation pathway.

Some technologies provide a similar service (e.g. wind and solar provide renewable electricity, BECCS and DAC provide carbon removals). To answer the question of which technology is 'better', we develop a method to explore the conditions under which one is preferred over the other. We measure the share of a service that is provided by each technology as a function of the relative cost between them: for example, the share of electricity from solar PV as a function of solar PV cost divided by onshore wind cost. Across several cases, we find that this relationship can be described by a logistic function in logarithmic space:

$$S_x = \frac{E_x}{E_x + E_y} = S_x^{min} + \frac{S_x^{max} - S_x^{min}}{1 + \exp\left(a + b \ln \frac{C_x}{C_y}\right)}$$

When considering two technologies, $x$ and $y$, the share of technology $x$ ($S_x$) is its output ($E_x$) as a proportion of total output ($E_x + E_y$). This is modelled by a logistic function which is bound to lower and upper levels based on the 2.5th percentile ($S_x^{min}$) and 97.5th percentile ($S_x^{max}$). This allows the curve to represent situations where a technology cannot supply all or none of a service (e.g. wind power cannot provide 100% of renewable electricity in 2030, as a non-zero capacity of solar power is still operating). The ratio of the capital cost of technologies $x$ and $y$ ($C_x/C_y$) is transformed into logarithmic space, so that technology $x$ being half or twice as expensive are equidistant from the central point, and this log-ratio is then transformed into a logistic function. Parameters $a$ and $b$ are fit through non-linear least-squares regression, where $a$ determines the position of the inflection point at which $S_x$= 0.5 (i.e. at what cost ratio are the technologies equally used) and $b$ describes the width of the cost window over which one technology takes share from the other.

To represent more complex competitions, we simplify to using the linear discriminant analysis (LDA) approach to supervised classification. Each of the 10,000 scenarios was classified according to whether one or more technologies met a defined condition, in the case of Figure 6 whether they supplied > 25% of total electricity. In the example with three classes, the discriminant scores from the LDA ($\delta_i$, $\delta_j$, $\delta_k$) were evaluated on a fine mesh (1000 x 1000) over the range of relative costs ($x$ and $y$). The boundaries of equal probability were extracted, for example where $\delta_i(x, y) = \delta_j(x, y)$, and summarised as a straight line with ordinary-least squares regression.

## Limitations

The national models used here adopt a single-node representation with sectoral demands largely exogenous in the aggregated variants, and they rely on time-series aggregation that necessarily compresses extremes. Both choices support computational feasibility given the need for 10,000 model runs, but may under-represent network constraints and operational stress conditions. The analysis also omits explicit constraints on technology diffusion beyond selected resource limits, and certain operational details (e.g., flexibility limits for firm generation) are simplified, which may affect the relative attractiveness of competing firm and flexible options in some futures. Finally, the probabilistic inputs are built from distributions (often Gaussian with specified percentiles) and do not fully encode potential correlations among technology costs or shared macroeconomic drivers (e.g. LFP and NMC battery costs will be partly correlated). These caveats do not undermine the central contribution around propagating wide uncertainty into pathway envelopes and tipping metrics, but they do highlight where key sensitivities and decision-relevant thresholds should be further explored.

## Data availability

The full data inputs to our models and full model outputs are available at: XXXXXXXXXX. The processed data used to generate all figures is included with this submission (see Data.xlsx).

## Code availability

The code developed to parameterise and run NESTOR-DE and NESTOR-GB, post-process and visualise results is available from the GitHub code repository at https://github.com/Gian-M/monte_carlo_workflow/. The underlying ETHOS.NESTOR model is instance of ETHOS.FINE which can be found at https://doi.org/10.21105/joss.06274 and https://github.com/fzj-iek3-vsa/fine.

# Acknowledgements

## Funding

This publication was supported by an ERA Fellowship – Green Hydrogen of the German Academic Exchange Service (DAAD). GM, TS and JW acknowledge support by the Helmholtz Association under the program "Energy System Design" and by the German Federal Ministry for Economic Affairs and Energy (BMWE) for funding this work within the SEDOS project (grant number: 03EI1040A).

## Author Contributions

Conceptualization: GM and IS
Data curation: GM and IS
Formal analysis: GM
Funding acquisition: JW, GM
Supervision: TS, JW, IS
Writing – original draft: GM and IS
Writing – review & editing: GM, TS, JW, IS

## Competing Interests

The authors declare no competing interests.

## Correspondence

Correspondence and requests for materials should be addressed to IS